\documentclass{article}
\usepackage{ltwol2e_ub}

\usepackage{psfig}

\def\Journal#1#2#3#4{{#1} {\bf #2}, #3 (#4)}

\def\NPB{{\em Nucl. Phys.} B}
\def\PLB{{\em Phys. Lett.}  B}

\def\PRD{{\em Phys. Rev.} D}
\def\ZPC{{\em Z. Phys.} C}

\def\be{\begin{equation}}
\def\ee{\end{equation}}
\def\bea{\begin{eqnarray}}
\def\eea{\end{eqnarray}}

\begin{document}

\title{ELECTROWEAK RADIATIVE CORRECTIONS TO $W$ BOSON PRODUCTION AT THE 
TEVATRON\footnote{}}

\author{U. Baur}

\address{Physics Department, State University of New York at Buffalo,
Buffalo, NY 14260}   


\twocolumn[\maketitle\abstracts{ We discuss the ${\cal
O}(\alpha)$ electroweak radiative corrections to $W$ boson production at 
the Tevatron and their effect on the $W$ boson mass extracted by 
experiment. The results of a new calculation of the ${\cal
O}(\alpha)$ corrections are presented and compared with those of a
previous calculation. We also briefly discuss the ${\cal O}(\alpha)$
corrections to $Z$ boson production at the Tevatron and two-photon
radiation in $W$ and $Z$ events. }]

\section{Introduction}
The Standard Model of electroweak interactions 
(SM)\footnote[0]{$^{\rm a}$Talk given at the XXIX International
Conference on High Energy Physics, Vancouver, B.C., Canada,
23~--~29~July 1998, to appear in the Proceedings} 
so far has met all experimental challenges and is now tested at the 
$0.1\%$ level~\cite{karlen}. However, there is little direct
experimental information on the mechanism  which generates the masses of
the weak gauge bosons. In the SM, spontaneous symmetry breaking is
responsible for mass generation. The existence of a Higgs boson
is a direct consequence of this mechanism. At present the negative
result of direct searches performed at LEP2 imposes a lower bound
of $M_H>89.8$~GeV~\cite{mcnam} on the Higgs boson mass. Indirect
information on the mass of the Higgs boson can be extracted from the 
$M_H$ dependence of radiative corrections to the $W$ boson mass, $M_W$,
and the effective weak mixing angle, $\sin^2\theta^{lept}_{eff}$.
Assuming the SM to be valid, a global $\chi^2$-fit to
all available electroweak precision data yields a 95\%
confidence level upper limit on $M_H$ of 280~GeV~\cite{karlen}. 

The current estimate of $M_H$ strongly depends~\cite{deg} on the world 
average for the weak mixing angle, $\sin^2\theta^{lept}_{eff}=0.23155 \pm
0.00019$~\cite{karlen}. It results from a combination of LEP and SLC
data which currently are not in good agreement~\cite{karlen}. Furthermore,
$\sin^2\theta^{lept}_{eff}$ is quite sensitive to the hadronic
contribution to $\alpha(M_Z^2)$, $\Delta\alpha_{\rm had}(M_Z^2)$. The
accuracy of $\Delta\alpha_{\rm had}(M_Z^2)$ has been the subject of a
number of publications during the last four years~\cite{hoecker}. Error 
estimates range between
$\delta(\Delta\alpha_{\rm had}(M_Z^2))=0.0007$~\cite{eid} and 
$\delta(\Delta\alpha_{\rm had}(M_Z^2))=0.00016$~\cite{dav}. A smaller
error for $\Delta\alpha_{\rm had}(M_Z^2)$ implies that
$\sin^2\theta^{lept}_{eff}$ receives more weight in the $M_H$ fit, {\it
i.e.} the discrepancy between the LEP and SLC data becomes a limiting
factor in the estimate of the Higgs boson mass from electroweak data. 

A more precise measurement of $M_W$ is, therefore, very important in 
order to extract more accurate information on $M_H$ from electroweak
data. Furthermore, in contrast to $\sin^2\theta^{lept}_{eff}$, the $W$ 
mass depends only mildly on $\Delta\alpha_{\rm
had}(M_Z^2)$~\cite{deg}. A more precise measurement of $M_W$ thus
automatically reduces the sensitivity of the extracted Higgs boson mass
to $\Delta\alpha_{\rm had}(M_Z^2)$. Currently, the $W$ boson mass is 
known to $\pm 0.06$~GeV~\cite{karlen} from direct measurements. A 
significant improvement in the $W$ mass uncertainty is expected in the
near future from measurements at LEP2~\cite{LEPWmass} and the 
Tevatron~\cite{Tev2000}. The ultimate precision
expected for $M_W$ from the combined LEP2 experiments is approximately 
40~MeV~\cite{LEPWmass}. At the Tevatron, integrated luminosities of
order 2~fb$^{-1}$ are foreseen for Run~II, and one
expects to measure the $W$ mass with a precision of approximately
40~MeV~\cite{Tev2000} per experiment and decay channel. 

In order to measure the $W$ boson mass with high
precision in a hadron collider environment, it is necessary to fully 
understand and control higher order QCD and electroweak (EW) corrections 
to $W$ production. The determination of the $W$ mass in a hadron 
collider environment
requires a simultaneous precision measurement of the $Z$ boson mass,
$M_Z$, and width, $\Gamma_Z$. When compared to the value measured at LEP, 
the two quantities help to accurately determine the energy scale and
resolution of the electromagnetic calorimeter, and to constrain the
muon momentum resolution~\cite{Tev2000}. It is therefore also necessary
to understand the higher order EW corrections to $Z$ boson production in 
hadronic collisions.

Recently, new and more accurate calculations of the ${\cal
O}(\alpha)$ EW corrections to $W$~\cite{BKW} and $Z$ boson production in 
hadronic collisions~\cite{BKS} became available. In a previous 
calculation, only the final state photonic corrections were correctly 
included~\cite{BK}. The sum of the soft and virtual parts was estimated
from the inclusive ${\cal O}(\alpha^2)$ $W\to\ell\nu(\gamma)$ and
$Z\to\ell^+\ell^-(\gamma)$ ($\ell=e,\,\mu$) 
width and the hard photon bremsstrahlung contribution. Initial state, 
interference, and weak contributions to the ${\cal O}(\alpha)$ 
corrections were ignored altogether. The unknown part
of the ${\cal O}(\alpha)$ EW corrections in
Ref.~[\ref{ber}], combined with effects of multiple
photon emission, have been estimated to
contribute a systematic uncertainty of $\delta M_W=
15-20$~MeV to the measurement of the $W$ mass~\cite{Tev2000}. 

In Section~2, we briefly describe the technical details of the calculation
of the ${\cal O}(\alpha)$ corrections to $W$ boson production presented
in Ref.~[\ref{wcalc}], and compare the results with those of
Ref.~[\ref{ber}]. In Section~3 we summarize the calculation of the
${\cal O}(\alpha)$ QED corrections to $Z$ boson production reported in
Ref.~[\ref{zet}], and in Section~4 some preliminary results of a new
calculation~\cite{BS} of two-photon radiation in $W$ and $Z$ production in
hadronic collisions are presented. 

\section{Electroweak Corrections to $W$ Boson Production at the Tevatron}
The calculation of the ${\cal O}(\alpha)$ corrections to $p\bar p\to
W\to\ell\nu$~\cite{BKW} is based on the full set of ${\cal
O}(\alpha^3)$ Feynman diagrams, and includes both initial and final 
state radiative corrections, as well as the contributions from their
interference. Final state
charged lepton mass effects are included in the following approximation. 
The lepton mass regularizes the collinear 
singularity associated with final state photon radiation. The associated
mass singular logarithms of the form $\ln(\hat s/m_\ell^2)$, where $\hat
s$ is the squared parton center of mass energy and $m_\ell$ is the
charged lepton mass, are included in our calculation, but the very small
terms of ${\cal O}(m_\ell^2/\hat s)$ are neglected. 

To perform the calculation, a Monte Carlo method for
next-to-leading-order (NLO) calculations similar to that described in 
Ref.~[\ref{mc}] was used. With the Monte Carlo method, it is easy to 
calculate a variety of observables simultaneously and to simulate detector 
response. Calculating the EW radiative corrections to $W$ boson
production, the problem arises how an unstable charged gauge boson can
be treated consistently in the framework of perturbation theory.
This problem has been studied in Ref.~[\ref{dor}] with 
particular emphasis on finding a gauge invariant decomposition of the 
EW ${\cal O}(\alpha)$ corrections into a QED-like and a modified weak part. 
In $W$ production, the Feynman diagrams which involve
a virtual photon do not represent a gauge invariant subset. 
In Ref.~[\ref{dor}] it was
demonstrated how gauge invariant contributions that contain the
infrared (IR) singular terms can be extracted from the virtual photonic 
corrections. These contributions can be combined with the also IR-singular 
real photon corrections in the soft photon region to form IR-finite 
gauge invariant QED-like contributions
corresponding to initial state, final state and interference
corrections. The IR finite remainder of the virtual photonic corrections 
and the pure weak one-loop corrections can be combined to separately
gauge invariant modified weak contributions to the $W$ boson
production and decay processes. 

The collinear singularities associated with initial state
photon radiation can be removed by universal collinear counter terms
generated by ``renormalizing'' the parton distribution
functions (PDF's)~\cite{spies}, in complete analogy to gluon emission in
QCD. In addition to the collinear counterterms, finite terms can be absorbed
into the PDF's, introducing a QED factorization scheme dependence. We
have carried out our calculation in the QED $DIS$ and QED $\overline{MS}$
scheme. In order to treat the ${\cal O}(\alpha)$ initial state QED-like
corrections to $W$ production in hadronic collisions in a
consistent way, QED corrections should be incorporated in the global 
fitting of the PDF's using the same factorization scheme which has been
employed to calculate the cross section. Current 
fits to the PDF's do not include QED corrections. A
study of the effect of QED corrections on the evolution of the parton 
distribution functions indicates~\cite{spies} that the modification 
of the PDF's is small. The 
missing QED corrections to the PDF introduce an uncertainty which,
however, is likely to be smaller than the present uncertainties on 
the parton distribution functions. 

\begin{figure}
\center
\psfig{figure=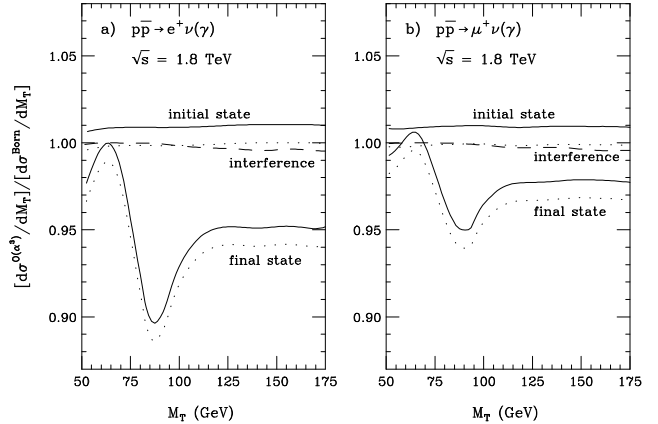,height=2.2in}
\caption{Ratio of the ${\cal O}(\alpha^3)$ and lowest order 
cross sections as a function of the transverse mass for a) $p\bar p\to
e^+\nu(\gamma)$ and b) $p\bar p\to \mu^+\nu(\gamma)$ at
$\protect{\sqrt{s}=1.8}$~TeV for various individual contributions. The 
upper (lower) solid lines show the result for the QED-like initial 
(final) state corrections. The upper (lower) dotted lines give the cross
section ratios if
both the QED-like and modified weak initial (final) state corrections are 
included. The dashed lines display the result if only the 
initial -- final state interference contributions are included.}
\label{fig:one}
\end{figure}

Since hadron collider detectors cannot directly 
detect the neutrinos produced in the leptonic $W$ boson decays,
$W\to\ell\nu$, and cannot measure the
longitudinal component of the recoil momentum, there is insufficient
information to reconstruct the invariant mass of the $W$ boson. 
Instead, the transverse mass ($M_T$) distribution of the $\ell\nu$ system is 
used to extract $M_W$. The various individual contributions to the
EW ${\cal O}(\alpha)$ corrections of the $M_T$ distribution are shown
in Fig.~\ref{fig:one}. 
To compute the cross section, we have used here the MRSA set of parton 
distribution functions~\cite{MRSA}. The detector acceptance is simulated 
by imposing the following
transverse momentum ($p_T$) and pseudo-rapidity ($\eta$) cuts:
\begin{equation}
p_T(\ell)>25~{\rm GeV,}\qquad |\eta(\ell)|<1.2, \qquad
\ell=e,\,\mu ,
\label{eq:lepcut}
\end{equation}
\begin{equation}
p\llap/_T>25~{\rm GeV.}
\label{eq:ptmisscut}
\end{equation}
These cuts approximately model the acceptance cuts used by the CDF and D\O\
collaborations in their $W$ mass analyses.
Uncertainties in the energy and momentum measurements of the charged leptons 
in the detector are simulated in the calculation by Gaussian smearing 
of the particle four-momentum vector using the specifications for the 
upgraded Run~II D\O\ detector~\cite{d0upgr}. 

The initial state QED-like 
contribution uniformly increases  the cross section by about 1\% for 
electron (Fig.~\ref{fig:one}a) and muon (Fig.~\ref{fig:one}b) final 
states. It is largely canceled by the modified weak initial state 
contribution. The interference contribution is very small. It decreases
the cross section by about $0.01\%$ for transverse masses below $M_W$, and
by up to $0.5\%$ for $M_T>M_W$. The final state QED-like contribution 
significantly changes 
the shape of the transverse mass distribution and reaches its maximum
effect in the region of the Jacobian peak, $M_T\approx M_W$. Since the 
final state QED-like contribution is proportional to $\ln(\hat
s/m_\ell^2)$, its size for muons is considerably
smaller than for electrons. As for the initial state,
the modified weak final state contribution reduces the cross section by 
about $1\%$, and has no effect on the shape of the transverse 
mass distribution. 

In Fig.~\ref{fig:one}, we have not taken into account realistic lepton
identification requirements. When these requirements are included, 
the mass 
singular logarithmic terms are eliminated in the electron case because 
the electron and photon momentum four vectors are combined for small 
opening angles where it is difficult to resolve the two
particles~\cite{BKW}. This significantly reduces the size of the EW 
corrections. On the other hand, in order 
to experimentally identify muons, the energy of the photon is required
to be smaller than a critical value if the $\mu-\gamma$ separation is 
small, and mass singular terms survive. Removing energetic
photons thus enhances the effect of the ${\cal O}(\alpha)$ corrections, 
and the effect of the EW corrections in the muon case is larger
than in the electron case once lepton identification requirements are
included.

As we have seen, final state bremsstrahlung has a non-negligible effect
on the shape of the $M_T$ distribution in the Jacobian peak region. It
is well known that EW corrections must be included when the
$W$ boson mass is extracted from data, otherwise the mass is 
shifted to a lower value. In the approximate treatment of
the electroweak corrections used so far by the Tevatron experiments,
only final state QED corrections are taken into account; initial state,
interference, and weak correction terms are ignored. Furthermore, the 
effect of the final state soft and virtual photonic
corrections is estimated from the inclusive ${\cal O}(\alpha^2)$ 
$W\to\ell\nu(\gamma)$ width and the hard photon 
bremsstrahlung contribution~\cite{BK}. When detector effects are
included, the approximate calculation leads to a shift of about
$-50$~MeV in the electron case, and approximately $-160$~MeV in the muon
case~\cite{Tev2000}.

\begin{figure}
\center
\psfig{figure=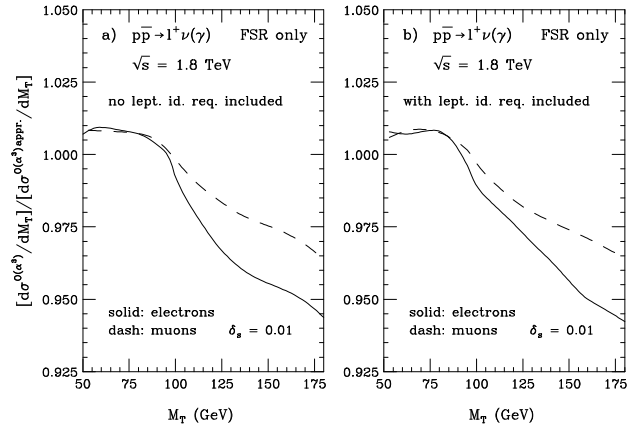,height=2.2in}
\caption{Ratio of the $M_T$ distributions obtained with the QED-like
final state correction part of Ref.~[\ref{wcalc}] to the one obtained using
the approximation of Ref.~[\ref{ber}] for
$p\bar p\to\ell^+\nu(\gamma)$ at $\protect{\sqrt{s}=1.8}$~TeV. }
\label{fig:two}
\end{figure}

Initial state and 
interference contributions do not change the shape of the $M_T$
distribution significantly (see Fig.~\ref{fig:one}) and therefore have
little effect on the
extracted mass. However, correctly incorporating the final state 
virtual and soft photonic corrections results in a non-negligible 
modification of the shape of the transverse mass distribution for
$M_T>M_W$. This is demonstrated in
Fig.~\ref{fig:two}, which shows the ratio of the $M_T$ distribution
obtained with the QED-like final state correction part of our
calculation to the one obtained using the approximation of
Ref.~[\ref{ber}]. 

The difference in the line shape of the $M_T$ distribution between the
${\cal O}(\alpha^3)$ calculation of Ref.~[\ref{wcalc}] and the approximation 
used so far occurs in a region which is important for both the
determination of the $W$ mass, and the direct measurement of the $W$
width. The precision which can be achieved in a measurement of $M_W$
using the transverse mass distribution strongly depends on how steeply
the $M_T$ distribution falls in the region $M_T\approx M_W$. Any change 
in the theoretical prediction of the line shape thus directly influences 
the $W$ mass measurement. From a maximum likelihood analysis the shift
in the measured $W$ mass due to the correct
treatment of the final state virtual and soft photonic corrections is
found to be $\Delta M_W\approx {\cal O}(10~{\rm MeV})$. This shift is 
much smaller than the present uncertainty for
$M_W$ from hadron collider experiments~\cite{Tev2000}. However,
for future precision experiments, a difference of ${\cal O}(10~{\rm
MeV})$ in the extracted value of $M_W$ can no longer be ignored, and the
complete ${\cal O}(\alpha^3)$ calculation should be used. 

\section{Electroweak Corrections to $Z$ Boson Production at the
Tevatron} 
The calculation of the ${\cal O}(\alpha)$ corrections to $Z$ boson
production~\cite{BKS} employs the same Monte Carlo method which was used 
in the $W$ case. The collinear
singularities originating from initial state photon radiation are
again removed by counter terms generated by renormalizing the PDF's. 
However, in contrast to $W$ production, the Feynman diagrams contributing to 
the ${\cal O}(\alpha)$ QED corrections can 
be separated into gauge invariant subsets corresponding to 
initial and final state corrections. Furthermore, the purely weak
corrections form a separately gauge invariant set of diagrams. The weak 
corrections are expected to be very small and are therefore
ignored in our calculation. Both $Z$ and photon exchange diagrams with all 
$\gamma - Z$ interference effects are incorporated. 

\begin{figure}
\centerline{
\psfig{figure=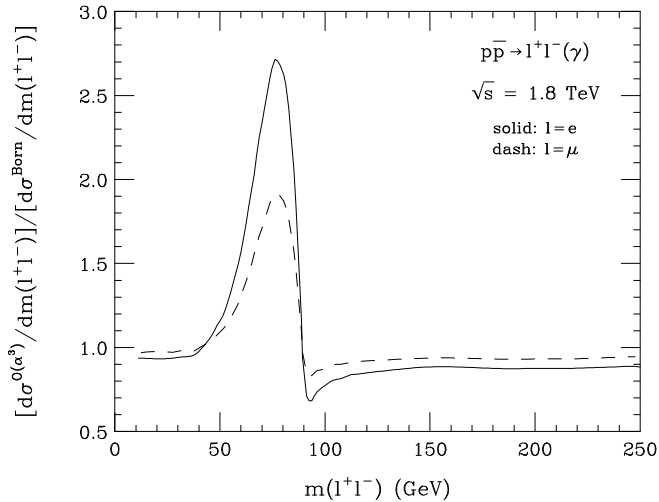,height=2.6in}}
\caption{Ratio of the \protect{${\cal O}(\alpha^3)$} and lowest order
differential cross sections as a function of the di-lepton invariant
mass for $p\bar p\to\ell^+\ell^-(\gamma)$ at $\protect{\sqrt{s}=1.8}$~TeV.}
\label{fig:three}
\end{figure}

In Fig.~\ref{fig:three} we display
the ratio of the ${\cal O}(\alpha^3)$ and the Born cross section as a 
function of the $\ell^+\ell^-$ invariant mass in $p\bar p\to
Z\to\ell^+\ell^-$. For
$40~{\rm GeV}<m(\ell^+\ell^-)<110$~GeV, the cross section ratio is seen
to vary rapidly. 
Below the $Z$ peak, QED corrections enhance the cross section by up to a 
factor~2.7 (1.9) for electrons (muons). The maximum enhancement of the 
cross section occurs at $m(\ell^+\ell^-)\approx 75$~GeV. At the $Z$ peak, the
differential cross section is reduced by about $30\%$ ($20\%$). For 
$m(\ell^+\ell^-)>130$~GeV, the ${\cal O}(\alpha)$ QED corrections 
uniformly reduce 
the differential cross section by about $12\%$ in the electron case, and 
$\approx 7\%$ in the muon case. No lepton identification requirements
are taken into account in Fig.~\ref{fig:three}. When these are included,
the maximum enhancement is reduced to a factor~1.6 for both electrons
and muons. As for the $W$, initial state corrections are uniform and
small. Final state radiative corrections dominate over the entire
di-lepton invariant mass range.

From Fig.~\ref{fig:three} it is clear that final state bremsstrahlung 
severely distorts the Breit-Wigner shape of the $Z$ resonance curve. As
a result, QED corrections must be included when the $Z$ boson mass is
extracted from data, otherwise the mass extracted is shifted to a lower
value. As in the $W$ case, in the approximate treatment of
the QED corrections to $Z$ boson production used so far by the Tevatron 
experiments, only final state corrections are taken into account, and the 
effects of soft and virtual corrections are estimated from the inclusive 
${\cal O}(\alpha^2)$ $Z\to\ell^+\ell^-(\gamma)$ width and the hard photon 
bremsstrahlung contribution~\cite{BK}. When detector effects are taken
into account, the approximate calculation leads to a shift of the $Z$
mass of about $-150$~MeV in the electron case, and approximately 
$-300$~MeV in the muon case~\cite{Tev2000}. The $Z$ boson mass extracted 
from our ${\cal O}(\alpha^3)$ $\ell^+\ell^-$ invariant mass distribution
is found to be about 10~MeV smaller than that obtained using the 
approximate calculation of Ref.~[\ref{ber}]. This translates into an
additional shift of several MeV in $M_W$ through the dependence of the 
energy scale and the momentum resolution on the $Z$ boson mass
measured. 

\section{Two Photon Radiation in $W$ and $Z$ Boson Production at the
Tevatron} 
The ${\cal O}(\alpha)$ EW
corrections have a significant effect on the $W$ and $Z$ masses extracted
by D\O\ and CDF. In particular, the large shift in the masses caused by
the emission of a photon from the final state lepton line raises the
question of how strongly multiple photon radiation influences the measured 
weak boson masses. At ${\cal O}(\alpha^n)$, $W$ or $Z$ decay with collinear 
emission of photons from a final state charged lepton gives rise to terms 
which are proportional to $(\alpha/\pi)^n\ln^n(M_V^2/m_\ell^2)$
($V=W,\,Z$) in $n$-photon exclusive rates.  

In order to find out how important multi photon radiation in $W$ and $Z$ 
production is for the measurement of $M_W$ at the Tevatron, it is
instructive to carry out a calculation of the two photon processes, $p\bar
p\to\ell\nu\gamma\gamma$ and $p\bar p\to\ell^+\ell^-\gamma\gamma$.
So far no calculation of these processes which 
is based on the full set of tree level ${\cal O}(\alpha^4)$ Feynman
diagrams, and which is valid for arbitrary lepton-photon opening
angles, has been carried out. For example, the calculation of
Ref.~[\ref{wgg}] assumes that $m_\ell=0$. A
non-zero $\Delta R_{\ell\gamma}$ cut, therefore, has to be imposed in
order to avoid the collinear singularities. The
Monte Carlo generator PHOTOS~\cite{photos}, on the other hand, treats
final state photon radiation in the leading-log approximation and thus 
leads to results which can only be trusted in the collinear region. 
PHOTOS ignores initial state photon radiation altogether. 

In order to correctly take into account the effects of two photon
radiation in $W$ and $Z$ production, a calculation which gives correct
results for small as well as large lepton-photon opening angles is
required. Here we report some preliminary results of such a calculation
which is presently carried out~\cite{BS}. Due to the collinear
singularities associated with photon radiation from the charged lepton
lines, there are many different peaks in the multidifferential cross 
section. For an accurate 
evaluation of the cross section we therefore use a multiconfiguration
Monte Carlo integration routine which automatically maps the peaks in 
the differential cross section to a uniform function according to the 
pole structure
of the contributing Feynman diagrams~\cite{stelz}. The matrix elements,
taking into account finite lepton masses, are calculated using the 
MADGRAPH package~\cite{madg}, which
automatically generates matrix elements in HELAS format~\cite{HELAS}.
In order to maintain electromagnetic gauge invariance for
$\ell\nu\gamma\gamma$ production in 
presence of finite $W$ width effects, the $W$ propagator and the
$WW\gamma$ and $WW\gamma\gamma$ vertex functions in the amplitudes
generated by MADGRAPH are modified, using the prescription of 
Ref.~[\ref{wgg}]. 

In Table~\ref{tab:one}, we display the fraction of $W\to e\nu$ and $Z\to
e^+e^-$ events (in percent) containing one or two photons at the
Tevatron as a function of the minimum photon transverse energy,
$E_T^{\rm min}$, for $E_T^{\rm min}\geq 0.1$~GeV, the approximate tower
threshold of the electromagnetic calorimeters of CDF and D\O. To
simulate detector response, we have imposed the following acceptance cuts:
\begin{equation}
p_T(e)>20~{\rm GeV,}~ |\eta(e)|<2.5,~{\rm and}~ |\eta(\gamma)|<3.6.
\end{equation}
In the $W$ case, we require in addition that 
\begin{equation}
p\llap/_T>20~{\rm GeV}
\end{equation}
and
\begin{equation}
65~{\rm GeV}<M_T(e+n\gamma;\nu)< 100~{\rm GeV},
\end{equation}
where $M_T(e+n\gamma;\nu)$ is the cluster transverse mass of the
$(e+n\gamma)\nu$ system ($n=0,1,2$). For $Z$ events we require 
\begin{equation}
m(e^+e^-)>20~{\rm GeV}
\end{equation}
and
\begin{equation}
75~{\rm GeV}<m(ee+n\gamma)<105~{\rm GeV},
\end{equation}
where $m(e^+e^-)$ ($m(ee+n\gamma)$) is the $e^+e^-$ ($ee+n\gamma$)
invariant mass.
 \begin{table}
\begin{center}
\caption{Fraction of $W\to e\nu$ and $Z\to e^+e^-$ events at the
Tevatron (in percent) containing one or two photons
with $E_T(\gamma)>E_T^{\rm min}$. 
Fractions are obtained by normalization with respect to the Born cross
section. The cuts imposed are described in the text. }
\label{tab:one}
\vskip 0.3cm
\begin{tabular}{|c|cc|} 
\hline 
\raisebox{0pt}[12pt][6pt]{$E_T^{\rm min}$ (GeV) } & 
\raisebox{0pt}[12pt][6pt]{$W\to e\nu\gamma$} & 
\raisebox{0pt}[12pt][6pt]{$W\to e\nu\gamma\gamma$} \\
\hline
\raisebox{0pt}[12pt][6pt]{0.1} & 
\raisebox{0pt}[0pt][0pt]{28} & 
\raisebox{0pt}[0pt][0pt]{4.0} \\
\raisebox{0pt}[0pt][0pt]{0.3} & 
\raisebox{0pt}[0pt][0pt]{21} & 
\raisebox{0pt}[0pt][0pt]{2.3} \\
\raisebox{0pt}[0pt][0pt]{1} & 
\raisebox{0pt}[0pt][0pt]{14} & 
\raisebox{0pt}[0pt][0pt]{0.9} \\
\raisebox{0pt}[0pt][0pt]{3} & 
\raisebox{0pt}[0pt][0pt]{7.8} & 
\raisebox{0pt}[0pt][0pt]{0.2}  \\
\hline
\raisebox{0pt}[12pt][6pt]{$E_T^{\rm min}$ (GeV) } & 
\raisebox{0pt}[12pt][6pt]{$Z\to e^+e^-\gamma$} & 
\raisebox{0pt}[12pt][6pt]{$Z\to e^+e^-\gamma\gamma$} \\
\hline
\raisebox{0pt}[12pt][6pt]{0.1} & 
\raisebox{0pt}[0pt][0pt]{54} & 
\raisebox{0pt}[0pt][0pt]{16} \\
\raisebox{0pt}[0pt][0pt]{0.3} & 
\raisebox{0pt}[0pt][0pt]{42} & 
\raisebox{0pt}[0pt][0pt]{9.4} \\
\raisebox{0pt}[0pt][0pt]{1} & 
\raisebox{0pt}[0pt][0pt]{27} & 
\raisebox{0pt}[0pt][0pt]{3.9} \\
\raisebox{0pt}[0pt][0pt]{3} & 
\raisebox{0pt}[0pt][0pt]{15} & 
\raisebox{0pt}[0pt][0pt]{1.2} \\
\hline
\end{tabular}
\end{center}
\end{table}
For muon final states, the fraction
of events containing one (two) photons is roughly a factor~2 (4) smaller 
than the results shown for $W\to e\nu$ and $Z\to e^+e^-$. 
Table~\ref{tab:one} demonstrates that a significant fraction of weak boson 
events contains two photons. Multiple photon bremsstrahlung thus is 
expected to have a non-negligible effect on the $W$ mass extracted 
from experiment. 

\section*{Acknowledgements}
This work has been supported by the National Science Foundation 
grant PHY9600770.

\end{document}